\documentstyle[twocolumn,aps]{revtex}
\begin{document}
\pagestyle{empty}
\title{On the Origin of Quantum Oscillations in the Mixed State of
Anisotropic Superconductor}
\author{Lev P. Gor'kov$^{1,2}$}
\address{$^1$National High Magnetic Field Laboratory, Florida State
University, Tallahassee, FL 32310}
\address{$^2$L.D. Landau Institute for Theoretical Physics, Russian Academy
of Sciences, 117334 Moscow, Russia}
\date{Received $~~~~~~~~~~~~~~$}

\maketitle

\begin{abstract}
For the vortex lattice in an anisotropic superconductor with
well-separated cores $(H_{c1}\ll B\ll H_{c2})$ it is shown that sizeable
de Haas-van Alphen oscillations are caused by the levels' crossing of
the energy threshold separating localized and extended states of
excitations moving in the average magnetic field, $B$.
\end{abstract}

\vspace{.15in}


\vspace{.25in}

Recent experiments [1,2] show that the de Haas-van Alphen (dHvA) effect
persists in the superconducting (SC) state for the magnetic field, $B$,
as low as $B\sim 0.3\div 0.4 H_{c2}$.  The effective SC-gap, $\Delta$,
at such fields is large enough to preclude motion of an electron along a
closed Larmour orbit with some radius, $r_L\sim v_F/\omega_c$,
exceedingly larger than the coherence length, $\xi_0$, and the
intervortex distance, $d$ [3].  The dHvA-signal is expected to weaken
exponentially as $\exp (-\Delta/\omega_c)$ [4].  In that which follows
we suggest a new mechanism for the quantum oscillations in the mixed
SC-state.

In normal state the dHvA-effect is brought about by levels crossing the
chemical potential, $\mu$, with the field variation.  The oscillations
are periodic in $B^{-1}$ because a minor field change, $\Delta B/B\sim
\omega_c/\mu$, is enough to push a level across $\mu$.

Electron- or hole-like character of SC-excitations depends on the extent
their energy exceeds the gap.  Even for a ``d-wave'' superconductor [5]
levels cannot cross the chemical potential.  In this sense, there is no
difference between a ``d-wave'' or any other anisotropic SC.

It is shown below that a new energy threshold takes over the role of the
chemical potential in the SC-state.  Consider, for example, an
anisotropic superconductor with a spectrum, $\varepsilon({\bf
p})=\sqrt{v^2_F(p-p_F)^2+|\Delta ({\bf p})|^2}$.  Assume, for
simplicity, that the gap, $\Delta({\bf p})$, has only one maximum,
$\Delta_{max}$, and one minimum, $\Delta_{min}$, along the Fermi Surface
(FS).  Excitations with $\varepsilon({\bf p})>\Delta_{max}$ have
itinerant behavior, while at $\Delta_{max}>\varepsilon({\bf
p})>\Delta_{min}$ this is only true for excitations with a proper
${\bf p}$.  The latter become localized in a magnetic field,
for the Lorentz force changes the ${\bf p}$-direction.  Excitations with
the energy larger than $\Delta_{max}$, may move along an extended
Larmour orbit.

To pose this phenomenon as a theoretical problem, consider the 
limit of well separated vortices $d\gg\xi_0 (H_{c1}\ll B\ll H_{c2})$.
The vortex cores occupying only a minor
fraction of the volume may be neglected.  A typical electron trajectory
would run across the ``bulk'', where the gap amplitude is saturated:
\begin{eqnarray}
\Delta({\bf p}, {\bf r}) \cong\Delta({\bf p})\exp (i\varphi({\bf r}))
\end{eqnarray}
The method [5] to treat the problem is based on averaging the Gor'kov
system of Green functions over quasiclassical trajectories [6] (all
notations below from ref. [5]).  The Green functions are presented in
terms of the position, $\varphi$, of an electron along the FS. The core
of the method is given by eqs. (23-28) [5].  The Gor'kov matrix being
diagonalized, the whole problem reduces to solving the following
Schr\"{o}dinger equation:
\begin{eqnarray}
-\omega_c^2y''+[\Delta^2(\varphi)-\omega_c\Delta'(\varphi)]y=E^2y
\end{eqnarray}
The term, $-\omega_c\Delta'(\varphi)$, plays no role and will be
omitted.  In (2) we have also left out the term $h(\varphi)$ of eq. (31)
[5].  The Doppler shift (31) [5] {\it is} essential for the magnitude of
the effect, and will be taken into account later.

The eigen functions are given by solutions of (2) satisfying the
periodicity condition $(\varphi\rightarrow\varphi +2\pi)$ for 
\begin{eqnarray}
y(\varphi)e^{-i\kappa\varphi}
\end{eqnarray}
where $\kappa =\bar{\mu}/\omega_c$ comes from presenting the chemical
potential in the form [5]:
\begin{eqnarray}
\mu=\omega_cN_o+\bar{\mu}
\end{eqnarray}
With $\mu$ being large, $N_0\gg 1$, and the specific $N_0$ does not
affect the pattern of periodic (in $B^{-1}$) oscillations of the
magnetization which depends on $\kappa$ in the interval (0,1).  Eq. (2)
and (3) become the problem of finding the band structure for a particle
moving in the periodic potential $\Delta^2(\varphi)$ with $\kappa$ as a
quasimomentum.

Re-write (2) in the form:
\begin{eqnarray}
-\omega_c^2y''+(\Delta^2(\varphi)-\Delta^2_{max})y=(E^2-\Delta^2_{max})y
\eqnum{2'}
\end{eqnarray}
and consider (2') first in the quasiclassical (WKB)-approximation
$(\omega_c\ll\Delta)$.  At $|E|\gg\Delta$ the periodicity of (3) leads
to the spectrum of free electrons in the magnetic field:
$E_n=\omega_cn+\bar{\mu}$.  At $E^2-\Delta^2_{max}<0$ the attractive
potential in (2') has many $(\Delta/\omega_c\gg 1)$ ``localized''
levels, (tunneling across the barrier is neglected).  The
boundary separating ``extended'' and ``localized'' states in the
WKB-sense lies at $\Delta_{max}$.  Introduce in
(2'):
\begin{eqnarray}
E^2-\Delta^2_{max}\simeq 2\Delta_{max}(-\varepsilon)
\end{eqnarray}
for $|E|$ close to $\Delta_{max}$.  The WKB-solutions are [7]:
\begin{eqnarray}
y_{\pm}(\varphi)=A(S'(\varphi))^{1/2}\exp [\pm
i\int^{\varphi}_0S'(\varphi)d\varphi] 
\end{eqnarray}
with 
\begin{eqnarray}
\omega_c S'(\varphi)=\left[
2\Delta_{max}(-\varepsilon)+\Delta^2_{max}-\Delta^2(\varphi)\right]^{1/2}
\end{eqnarray}
($A$ is the normalization coefficient).  The
BCS-factors, $u(\varphi)$ and $v(\varphi)$, in (23) [5] are to be
normalized together: $\overline{|u^2|}+\overline{|v^2|}=1$, (the
bar in $\overline{(\ldots)}$ means the normalization integral:
$(2\pi)^{-1}\int^{2\pi}_0(\ldots)d\varphi$).  Two auxiliary expressions
which follow from eqs. (26-28) [5]:
\begin{eqnarray}
\overline{|u^2|}=(1/2)\left\{\overline{|y|^2}+(i\omega_c/2E)
\overline{(y^{\ast}y'-yy^{\ast '})}\right\} \nonumber \\
\overline{|v^2|}=(1/2)\left\{\overline{|y|^2}-(i\omega_c/2E)
\overline{(y^{\ast}y'-yy^{\ast '})}\right\}
\end{eqnarray}
immediately show that $\overline{|y|^2}=1$.

Expression (6) [5], containing oscillatory effects 
\begin{eqnarray}
M=-\frac{\mu e}{\pi c}\sum_{\lambda}\overline{|u_{\lambda}(\varphi)|^2}
\end{eqnarray}
($\lambda$ enumerates the eigenvalues, factor 2 added to (6) [5],
accounts for spins, and $n(E_{\lambda})\equiv 1$ at $E_{\lambda}<0$ and
$T=0$), becomes an integral over $\lambda$ with the use of
the Poisson  formula:
\begin{eqnarray}
\sum_{-\infty}^{+\infty}\delta(\lambda -n)=\sum_{k=-\infty}^{+\infty}
e^{2\pi iK\lambda}
\end{eqnarray}
Integration by parts transforms $M_{osc}$ into [5]:
\begin{eqnarray}
M_{osc}=\frac{i\mu e}{2\pi^2c}\sum_K
\frac{1}{K}\int_{-\infty}^{+\infty}e^{2i\pi K\lambda}
\frac{d}{d\lambda}\left(\overline{|u_{\lambda}(\varphi)|^2}\right)
d\lambda
\eqnum{10'}
\end{eqnarray}
Although integration over $\lambda$ acquires the meaning only after the
connection between $\lambda$ and the energy is established, the
threshold separating ``localized'' and ``extended'' (in the WKB-sense
(6,7)) states is already seen in eq. (10'): for ``localized'' states,
$|E|<\Delta_{max}$, the wave functions are real, and from (8)
$\overline{|u_{\lambda}^2|}=1/2$.  For ``extended'' states (6) we have:
\begin{eqnarray}
\overline{|u_{\lambda}|^2}=\frac{1}{2}\left\{
1\pm\frac{\omega_c}{E_{\lambda}}|A_{\lambda}|^2\right\}
\end{eqnarray}
with $|A_{\lambda}|^2$-a function of energy (see (6)).  The derivative
in (10') thus {\it eliminates states below $\Delta_{max}$} with
$\overline{|u_{\lambda}|^2}=1/2$ being energy independent.

Returning to summation over $\lambda$, eqs. (10) and
(10'), we need to construct such the function:
\begin{eqnarray}
\lambda (E)=\Phi(E)/2\pi
\end{eqnarray}
that the provision:
\begin{eqnarray}
\Phi(E_n)=2\pi n \eqnum{12'}
\end{eqnarray}
would enumerate all energy levels in the consecutive order.   In the
WKB-approach $\Phi(E)$ is given by:
\begin{eqnarray}
S(2\pi,-\varepsilon)=(1/\omega_c)\int^{2\pi}_0 S'(\varphi
,-\varepsilon)d\varphi
-2\pi\kappa
\end{eqnarray}
which at large energies matches the Landau free electron spectrum.  The
approach falls short near $(-\varepsilon)=0$.

Choose $\Delta^2(\varphi)$ near $\Delta_{max}$ as 
\begin{eqnarray}
\Delta^2(\varphi)=\Delta^2_{max}(1-a\varphi^2)
\end{eqnarray}
Expanding (13) in $(-\varepsilon)>0$, one obtains:
\begin{eqnarray}
S(2\pi,-\varepsilon)\simeq S(2\pi,0)-(l/2) \ln (l/\wedge)
\end{eqnarray}
with the useful notation in (14):
\begin{eqnarray}
l=(-2\varepsilon) /\omega_c a^{1/2}~;~~\wedge
=(a^{1/2}\Delta_{max}/\omega_c)\gg 1
\end{eqnarray}
Similarly, the factor $|A|^2$ in (6) and (11) at small $l$ is
proportional to:
\begin{eqnarray}
|A|^2\propto \left[ \ln(l/\wedge)\right]^{-1} \eqnum{15'}
\end{eqnarray}

Because of the singularity (15) the ``numbering'' function $\Phi(E)$
cannot be comprised of the two WKB-branches, the one that is given by
(13) (at $(-\varepsilon)>0$), and the other which counts 
``localized'' states $((-\varepsilon)>0)$. 

Note that far away from $\varphi =0$ the WKB-solution
\begin{eqnarray}
y(\varphi)=ay_+(\varphi)+by_-(\varphi)
\end{eqnarray}
is still correct.  With the use of (5,14,16), eq. (2') can be solved 
near $\varphi =0$ in terms of the parabolic cylinder
functions.  It establishes the matrix relation between coefficients
$(a,b)$ in (17) on the R.H.S. of $\varphi=0_+$, and the other set
$(a',b')$ on its L.H.S.,  $\varphi =0_-$:
\begin{eqnarray}
\left(\begin{array}{c} a'\\ b'
\end{array}\right)=\left(\begin{array}{cc} \alpha & \beta \\
\beta^{\ast} & \alpha^{\ast} \end{array}\right) \left( \begin{array}{c}
a\\ b \end{array}\right)
\end{eqnarray}
Beginning at $\varphi =0_+$, moving along with eq. (17) toward $(2\pi)_-$
and using (18), the periodicity condition for
(3) provides the equation:
\begin{eqnarray}
R(l)\equiv |\alpha |(e^{i\tilde{S}}+e^{-i\tilde{S}})=2\cos 2\pi\kappa
\end{eqnarray}
Abbreviations in eq. (19) are:
\begin{eqnarray}
\tilde{S}=S(2\pi,-\varepsilon)-\theta,~\alpha =|\alpha |\exp(i\theta)
\eqnum{19'}
\end{eqnarray}
Of the two solutions of eq. (19) we chose
\begin{eqnarray}
e^{i\tilde{S}}=\frac{1}{|\alpha |}\left\{ \cos 2\pi\kappa
+i\sqrt{|\alpha |^2-\cos^22\pi\kappa}\right\}\equiv\rho(l)
\end{eqnarray}
because at $(-\varepsilon)$, i.e. $l$, large and positive $|\alpha
|\Rightarrow 1$, $\theta\Rightarrow 0$, and we return in this limit to
(13).  On the real axis of $l$
the function
\begin{eqnarray}
\Phi(l)=S(2\pi,l)-\theta(l)-\frac{1}{i}\ln \rho(l)
\end{eqnarray}
is positive, with $d\Phi /dl >0$, and matches asymptotically (at
$l\rightarrow\infty$) the free electron spectrum.  With the help of (12)
and (21) the oscillatory magnetization is 
\begin{eqnarray}
M_{osc}=\frac{i\mu e}{2\pi^2c}\sum_K
\frac{1}{K}\int_{-\infty}^{+\infty}
e^{iK\Phi(l)}\frac{d}{dl}\left(\overline{|u_l(\varphi)|^2}\right)dl 
\end{eqnarray}

$\Phi(l)$ being continued analytically into the complex $l$-plane, the
 integration may be shifted into the upper half-plane
(from (11, 15') one has the behavior of
$d(\overline{|u_l|^2})/dl$ at large $|l|$).  Consider singularities in
(22).  Thus from the expression for $|\alpha |$
[8]:
\begin{eqnarray}
|\alpha |=(1+e^{-\pi l})^{1/2}
\end{eqnarray}
we conclude that branching points in (23) lie at 
\begin{eqnarray}
l_m=\pm (2m+1)i
\end{eqnarray}
This is also true for $\theta(l)$ (See $\tilde{S}(2\pi,l)$ below).  The
definition of $\rho (l)$ together with (23) for $|\alpha |$, leads to
the square root singularities at
\begin{eqnarray}
l_m'=\pm (2m+1)i-\frac{1}{\pi}\ln (\sin^22\pi\kappa)\equiv l_m+l_0
\eqnum{24'}
\end{eqnarray}
For $\Phi(l)$ to be analytical in a strip at the real axis, the
branch-cuts in the $l$-plane, caused by singularities (24, 24'), must be
chosen parallel to the imaginary axis.

The integral in (22) may be bent to the
contours, $C_1$ and $C_2$, each encircling the branch-cuts  (24)
and (24') in the upper half-plane.  The
non-analytical terms of eq. (15) at $|\lambda |\sim 1$ are now absent in
$\tilde{S}(2\pi,l) (c\sim 1)$:
\begin{eqnarray}
\tilde{S}(2\pi,l)\simeq S(2\pi,0)+ \frac{1}{2} l[\ln(\wedge\cdot c)]-
\nonumber \\
-\frac{1}{2i}\ln\left[
\Gamma(\frac{1}{2}+\frac{i l}{2})/\Gamma(\frac{1}{2}-\frac{i
l}{2})\right]  
\end{eqnarray}
Both integrals (along $C_1$ and $C_2$) rapidly converge.

It is necessary to normalize $\overline{|u_l(\varphi)|^2}$ with the
accuracy better than that given by the WKB-approximation in (6), (11).
Fortunately the properties of the Bloch functions in a one-dimensional
periodic potential are well-studied.  With the help of eq. (4.18) in
ref. [9] and our eqs. (8) we derive:
\begin{eqnarray}
\overline{|u_l(\varphi)|^2}=\frac{1}{2}-\pi a^{1/2}\sin 2\pi\kappa
\left[\frac{dR}{dl}\right]^{-1}
\end{eqnarray}
(Here $R(l)$ is the R.H.S. of (19)).  After differentiation on $l$ in
(22) had eliminated $(1/2)$ in (26),  one may use for the rest the rapid
convergence of the integrals to integrate back by
parts in (22).  Single terms under the sum  symbol (22) become:
\begin{eqnarray}
I_K=\frac{i\pi a^{1/2}\sin 2\pi\kappa}{2}\int \frac{\exp
(iK\Phi(l))dl}{(\sin^22\pi\kappa+e^{-\pi l})^{1/2}}
\end{eqnarray}
with integrals running along $C_1$, $C_2$.  Assume that $\ln\wedge\gg 1$
in (25).  First, the term in (22) with $K=1$ prevails.  In addition, as
seen from (25), it is enough to consider the nearest singularities with
$m=0$ in (24, 24').  Defining the branches of the square roots in (27)
properly, from (27) one obtains contributions to $I_1$ from the two
contour integrals, over $C_1$ and $C_2$, correspondingly:
\begin{eqnarray}
I_1(C_1)=-\frac{(2\pi)^{3/2}a^{1/2}tg2\pi\kappa
}{(\wedge c)^{1/2}(\ln\wedge c)^{1/2}}
e^{iS(2\pi,0)+\frac{i\pi}{4}} 
\eqnum{28'}
\\
I_1(C_2)=\frac{\pi a^{1/2}\sin 2\pi\kappa e^{\frac{i\pi}{4}}}{(\wedge
c)^{1/2}(\ln\wedge c)^{1/2}}\left[
\frac{\Gamma(1-\frac{il_0}{2})}{\Gamma(1+\frac{il_0}{2})}\right]^{1/2}\cdot
\nonumber \\
\cdot e^{iS(2\pi,0) + \frac{il_o}{2}\ln(\wedge c)} \eqnum{28''}
\end{eqnarray}
with $l_0$ from (24').  ($I_1$ has the form (28', 28'') at $l_0$
not too small $(l_0\stackrel{>}{\sim}(\ln\wedge c)^{-1})$.  Otherwise
the two contours $C_1$ and $C_2$ start to merge.  Also, if $l_0$ becomes
large $(\sin^22\pi\kappa\rightarrow 0)$, expression (25) for
$\tilde{S}(2\pi,l)$ being correct at $|l|\sim 1$, ceases to be
applicable).

Expressions (28) explicitly present the periodic (in $B^{-1}$)
oscillations in the magnetization as $\kappa$ varies in the interval
(0,1).  It is notable that the amplitude is of the order of
$(\omega_c/\Delta_{max})^{1/2}$, i.e., is not exponentially small.  Both
(28') and (28'') lead to the large content of the higher harmonics.
In principle, $M_{osc}$ could be measured directly as a function of
small changes in $B$.  $I_1(C_1)$ discloses a rather regular behavior in
$\Delta B$ (i.e., $\kappa$), while  $I_1(C_2)$ rapidly
becomes chaotic due to the phase factor in (28''), $\exp
\left[ i(l_0/2)\ln(\wedge c)\right]$, contributing into higher
harmonics.  At the Fourier analysis of the dHvA-signal a few
first harmonics are expected to be seen with the intensity of the
order of
\begin{eqnarray}
(\omega_c/\Delta_{max})^{1/2} \eqnum{29}
\end{eqnarray}

Unfortunately,  (29) does not take into account scattering
of electrons on the flux lines.  The term $h(\varphi)$ of eq. (31) [5],
if included, adds to the potential of eq. (2'):
\begin{eqnarray}
2\Delta_{max} h(\varphi) \eqnum{30}
\end{eqnarray}
Even though $h(\varphi)\sim v_F/d$ is small compared with
$\Delta_{max}$, (30) drastically distorts the potential {\it near} $\varphi
=0$.  It is a {\it local} $\tilde{h}_{max}$ in the vicinity of the maximum in
$\Delta(\varphi)$ which now sets in the energy threshold between
``localized'' and ``extended'' states.  Note that although $h(\varphi)$
is rather irregular (for a given trajectory) and does change typically
on the scale of $\delta\varphi\sim (d\omega_c/v_F)$, its local maxima
produce potential barriers in (30) which remain impenetrable in the
quasiclassical sense.  The above analysis of $M_{osc}$ can be performed
in exactly the same manner as above around $\tilde{h}_{max}$.  There is
a change in the scale (29), because the curvature, $a$, near a
maximum in (30) is much higher than in (14).  Without
going into details, we comment that this only increases the effect,
because the potential $h(\varphi)$ near $\tilde{h}_{max}$ comprises a
much sharper barrier, if compared with eq. (14).

The major destroying effect comes from the phase factor in (25),
$S(2\pi,0)$.  At $h(\varphi)\neq 0$, $S(2\pi,0)$ may be expanded in
$\delta h(\varphi)=h(\varphi)-\tilde{h}_{max}$.  (Now
$(-\varepsilon)\Rightarrow E-\Delta_{max}-\tilde{h}_{max}$).
The fluctuating part, $\delta S(2\pi,0)$, is
\begin{eqnarray}
\delta S(2\pi,0)=-(\Delta_{max}/\omega_c)\int_0^{2\pi}\delta
h(\varphi)\left(\Delta^2_{max}-\Delta^2(\varphi)\right)^{-1/2}d\varphi
\eqnum{31}
\end{eqnarray}
Since $\langle\delta h(\varphi)\rangle$ (the average over all 
trajectories) is obviously zero, fluctuations in $\exp(i\delta
S(2\pi,0))$ lead, as in [5], to an effective Dingle factor of the form:
\begin{eqnarray}
\exp(-v_F/d\omega_c)\Rightarrow\exp(-(\Delta/\omega_c)(\xi_0/d))
\eqnum{32}
\end{eqnarray}
The exponent in (32) provides much more favorable conditions for 
observation of the dHvA-effect than previous results [4].

To conclude, in the developed mixed state of an anisotropic
superconductor there exists the energy threshold sorting
excitations into two categories: the localized and extended ones.
Crossing this threshold by the excitations' levels at the change of the
magnetic field comprises the new mechanism for quantum oscillations.
Scattering on the flux lines reduces the dHvA-effect.  Nevertheless, the
effect remans  bigger than anticipated.

The work was supported by the NHMFL through NSF cooperative agreement
No. DMR-9016241 and the State of Florida.

\vspace{.25in}

\noindent {\bf References}

\vspace{.25in}

\noindent [1] T. Terashima {\it et al.}, Phys. Rev. {\bf B 56}, 5120 (1997).

\noindent [2] Y. Onuki, in {\it Proceedings of the Conference SCES '98},
\\ \indent $~~$Paris, 1998 (to
be published).

\noindent [3] A.A. Abrikosov, Sov. Phys. JETP {\bf 5}, 1174 (1957).

\noindent [4] K. Maki, Phys. Rev. {\bf B 44}, 2861 (1991), M.J. \\
\indent $~~$Stephen,
Phys. Rev. {\bf B 45}, 5481 (1992).

\noindent [5] L.P. Gor'kov and J.R. Schrieffer, Phys. Rev. Lett. \\
\indent $~~${\bf
80}, 3360 (1998).

\noindent [6] E.A. Shapoval, Sov. Phys. JETP {\bf 20}, 675 (1965).

\noindent [7] L.D. Landau and E.M. Lifshitz, {\it Quantum}\\ \indent
$~~${\it  Mechanics:
nonrelativistic theory} (Pergammon \\ \indent $~~$Press, New York, 1977).

\noindent [8] To the best of the author's knowledge, the complete \\ \indent
$~~$matrix in (17) was not published, although \\ \indent
$~~$transmission/ reflection processes for a parabolic \\ \indent
$~~$barrier have been studied (see in ref. [7]). The author \\ \indent
$~~$thanks V.  Pokrovsky for discussion of references.

\noindent [9] W. Kohn, Phys. Rev. {\bf 115}, 809 (1959).

\end{document}